\documentclass{elsart}
\usepackage{amsmath}
\usepackage{amssymb}
\usepackage{graphicx}
\journal{Physics Letters A}

\begin{document}
\begin{frontmatter}

\title{Exact solutions for magnetic annihilation in curvilinear geometry\thanksref{talk}}
\thanks[talk]{Article presented at the 7th Plasma Easter Meeting on the Nonlinear Dynamics of Fusion Plasmas, Turin, 3-5 April 2002.}
\author[Ruhruni]{E. Tassi\corauthref{cor}}\ead{tassi@tp4.ruhr-uni-bochum.de}, \author{V.S. Titov} and \author {G. Hornig}
\corauth[cor]{Corresponding author.}

\address{Theoretische Physik IV, Ruhr-Universit\"at Bochum, 44780 Bochum, Germany}

\address[Ruhruni]{Theoretische Physik IV, Ruhr-Universit\"at Bochum, 44780 Bochum, Germany, Telephone number: + 49 234 3223458, Fax number: + 49 234 32 14177}  

\begin{abstract}
New exact solutions of the steady and incompressible 2D MHD equations in polar coordinates are presented. The solutions describe the process of reconnective magnetic annihilation in a curved current layer. They are particularly interesting for modeling  magnetic reconnection in solar flares caused by the interaction of three photospheric magnetic sources.
\end{abstract}

\end{frontmatter}

\section{Introduction}
Magnetic reconnection is a fundamental process in many areas of plasma physics. In particular it is proved to play a key role in active phenomena such as solar flares and geomagnetic substorms \cite{PrFbs00,Bi00}. Since the pioneering works of Parker \cite{Pa57}, Sweet \cite{Sw58} and Petschek \cite{Pe64}, several exact solutions describing reconnective annihilation in two dimensions in Cartesian coordinates were discovered e.g. by Sonnerup and Priest \cite{So75}, Craig and Henton \cite{Cr95} and Priest et al. \cite{Pr00}. The aim of this paper is to present similar solutions for a curvilinear geometry in a form that makes it possible to model a class of solar flares.
    
\section{Basic equations}
We consider stationary and incompressible plasma flows with uniform density and resistivity which are described by the equation of motion 
\begin{equation}    \label{e:mot}
({\vec v}\cdot\nabla){\vec v}=-\nabla p +{{(\nabla\times{\vec B}}})\times{\vec B}
\end{equation}
and by the Ohm's law 
\begin{equation}  \label{e:Ohm}
{\vec E}+{\vec v}\times{\vec B}=\eta\nabla\times{\vec B},
\end{equation}
where the velocity ${\vec v}$ and magnetic field ${\vec B}$ satisfy
\begin{equation} \label{e:divfree}
\begin{split}
&\nabla\cdot{\vec v}=0,\\
&\nabla\cdot{\vec B}=0.\\
\end{split}
\end{equation}
The equations (\ref{e:mot}) and (\ref{e:Ohm}) are written in a dimensionless form. The distances have been rescaled with respect to a characteristic length $L_{\mathrm{e}}$. The magnetic field and the plasma velocity have been normalized respectively to $B_{\mathrm{e}}$ and $v_{\mathrm{Ae}}$ which are characteristic values of the magnetic field and of the Alfv\a'en velocity. The nondimensional resistivity $\eta$ corresponds to the inverse magnetic Reynolds number.
The current density ${\vec j}$ is determined by the Ampere's law
\begin{equation}   \label{e:Ampe}
{\vec j}=\nabla \times {\vec B}.
\end{equation}
Assume that the velocity and magnetic fields lie in one plane and do not depend on the $z$-coordinate perpendicular to this plane. Then the electric field ${\vec E}$ is uniform and parallel to the $z$-axis. In the plane where the magnetic and velocity field lie we use polar coordinates  $(r,\theta)$ related to the Cartesian coordinates $(x,y)$ as follows
\begin{equation}  \label{e:polcoord}
x=r\sin \theta, \qquad y=r \cos\theta -d,
\end{equation}
where $d>0$ so that the pole is below the line $y=0$. The above mentioned $L_{\mathrm{e}}$ is the distance from the origin of the coordinate system. Further we restrict our consideration to the domain $(y>0, r<1)$ in the corona, since the subphotospheric flows are described by other equations which at least have to take into account the gravity. Due to (\ref{e:divfree}) the velocity and magnetic fields can be expressed in terms of stream and flux functions $\psi$ and $A$, respectively. In polar coordinates this yields
\begin{equation}
(v_{\mathrm{r}}, v_{\theta})=\left(\frac{1}{r}{\frac{\partial \psi}{\partial \theta}}, -{\frac{\partial \psi}{\partial r}}\right), \qquad (B_r, B_{\theta})=\left(\frac{1}{r}{\frac{\partial A}{\partial \theta}}, -{\frac{\partial A}{\partial r}}\right).
\end{equation}
Eqs. (\ref{e:mot}) and (\ref{e:Ohm}) rewritten in terms of $\psi$ and $A$, have, respectively, the following form:
\begin{equation}    \label{e:mot2}
[\psi,{\nabla}^2 \psi]=[A,{\nabla}^2 A],
\end{equation}
\begin{equation}    \label{e:Ohm2}
E+\frac{1}{r}[\psi,A]=-\eta{\nabla}^2 A.
\end{equation} 
Here the Poisson brackets are defined as
\begin{equation}
[f,g]={\frac{\partial f}{\partial r}}{\frac{\partial g}{\partial \theta}}-{\frac{\partial g}{\partial r}}{\frac{\partial f}{\partial \theta}}.
\end{equation}

\section{Form of the solutions}
We seek solutions in the form
\begin{equation}  \label{e:ansA}
A(r,\theta)=A_1(r)\theta+A_0(r),
\end{equation}
\begin{equation}  \label{e:ansP}
\psi(r,\theta)=\psi_1(r)\theta+\psi_0(r),
\end{equation}
where $A_1$, $\psi_1$, $A_0$ and $\psi_0$ are unknown functions of $r$.
Then the radial and azimuthal components of the magnetic and velocity fields are
 \begin{equation} \label{e:Bv}
 \begin{split}
  &B_{\mathrm{r}}(r)=\frac{A_1}{r}, \qquad B_{\theta}(r,\theta)=-{A_1}^{\prime}\theta-{A_0}^{\prime},\\
  & v_{\mathrm{r}}(r)=\frac{\psi_1}{r}, \qquad v_{\theta}(r,\theta)=-{\psi_1}^{\prime}\theta-{\psi_0}^{\prime},\\
 \end{split}
 \end{equation}
in which the symbol $\prime$ indicates the derivative with respect to $r$.\\
By substituting the ansatz (\ref{e:ansA}) and (\ref{e:ansP}) into (\ref{e:mot2}) and (\ref{e:Ohm2}) one obtains expressions depending linearly on $\theta$. This means that in each of these expressions the coefficient of $\theta$ and the sum of the terms not depending on $\theta$ have to vanish. This yields a set of four ordinary differential equations which splits into two subsystems. The first of them is nonlinear and includes only $A_1$ and $\psi_1$, so that  
\begin{equation}  \label{e:i4}
\frac{{\psi_1}^{'}}{r}{\left(r{\psi_1}^{'}\right)}^{'}
-\psi_1{\left[{\frac{1}{r}}{(r{\psi_1}')}^{'}\right]}^{'}=
{\frac{{A_1}'}{r}}{\left({r{A_1}'}\right)}^{'}-A_1{\left[{\frac{1}{r}}{(r{A_1}')}^{'}\right]}^{'},
\end{equation}
\begin{equation}  \label{e:i1}
  \psi_1 'A_1-\psi_1 A_1 '+\eta ({A_1}^{\prime}+ r {A_1}^{\prime\prime})=0.
\end{equation}
The second subsystem is linear in $A_0$ and $\psi_0$ and it has the form:
\begin{equation}  \label{e:i3}
{\frac{{\psi_0}'}{r}}{\left({r{\psi_1}'}\right)}'-\psi_1{\left[\frac{1}{r}{(r{\psi_0}')}'\right]}^{'}={\frac{{A_0}'}{r}}{\left({r{A_1}'}\right)}^{'}-A_1{\left[\frac{1}{r}{(r{A_0}')}^{'}\right]}^{'},
\end{equation}
\begin{equation}  \label{e:i2}
E+\frac{1}{r}[\psi_0 'A_1-\psi_1 A_0 '+\eta ({A_0}^{\prime}+ r {A_0}^{\prime\prime})]=0.
\end{equation}
Since the obtained set consists of four equations for four unknowns, the assumed ansatz is compatible with equations (\ref{e:mot2}) and (\ref{e:Ohm2}).\\ 

\section{Ideal solutions}
Let us consider first the case of ideal MHD, that is when $\eta=0$. In this limit it is easy to see from eq. (\ref{e:i1}) that $\psi_1$ must be proportional to $A_1$, that is
\begin{equation} \label{e:prop}
\psi_1=\alpha A_1,
\end{equation}
where $\alpha$ is an arbitrary constant. By using this result in (\ref{e:i4}) we obtain the equation
\begin{equation}  \label{e:i4sec}
{A_1}^{\prime\prime}+\frac{{A_1}^{\prime}}{r}\pm{\lambda}^2 A_1=0,
\end{equation} 
where $\lambda=-\left.\left({B_{\mathrm{r}}}^{-1} \partial j/ \partial \theta \right)\right\vert_{r=1}$.
Let us consider the case of $\alpha\neq \pm 1$. If $\lambda=0$, so that the current density does not depend on $\theta$, the solutions of the system for vanishing resistivity are given by
\begin{equation} \label{e:A1p1}
\psi_1=c_1 \ln r +c_2, \qquad A_1=\frac{1}{\alpha}(c_1\ln r +c_2),
\end{equation}
\begin{equation} \label{e:A0pgen}
 {A_0}^{\prime}=\frac{\alpha}{{{\alpha}^2-1}}{\frac{Er}{{c_1 \ln r +c_2}}}+\frac{a}{\alpha}r+\frac{b}{{\alpha r}}, \qquad  {\psi_0}^{\prime}=\frac{1}{{{\alpha}^2-1}}{\frac{Er}{{c_1\ln r +c_2}}}+ar+\frac{b}{r},
 \end{equation}
where $c_1$, $c_2$, $a$ and $b$ are arbitrary constants. From (\ref{e:A0pgen}), (\ref{e:Bv}) and (\ref{e:Ampe}), it is easy to see that in the presence of a non-vanishing electric field $E$ the $\theta$-components of the magnetic and velocity fields as well as the current density have a singularity at the radius $r=\exp(-c_2/c_1)\equiv r_{\mathrm{c}}$, which is further on called the critical radius. In ideal MHD this is an indication of reconnection. The singularities disappear at $E=0$ and in this case the arc $r=r_{\mathrm{c}}$ turns into the separatrix line emanating from the null point $(r_{\mathrm{c}},0)$. Thus, the electric field is responsible for the appearance of the singularities at the separatrix: in our solution it drives the shearing flows across the other separatrix line, which is exactly the situation where the magnetic flux has to pile up at the separatrix aligned with the shear \cite{Pr96}. The first example of such a solution in a more simple geometry has been discovered by Craig and Henton \cite{Cr95}, which showed also that the corresponding singularity is resolved by resistivity. It is shown below that this is valid for our solution as well.

\section{Resistive solutions}
To resolve the above singularity we adopt the method of matched asymptotic expansions by analogy with the work of Priest et al. \cite{Pr00}, where it is used for solving a similar problem in Cartesian rather than cylindrical geometry. One can use such a method, because the dimensionless resistivity $\eta$ is very small in the solar corona as well as in many other astrophysical and laboratory plasmas. To solve the problem in this case the domain is separated in two different regions, a narrow layer enclosing the singularity and the rest of the plane. In each of the regions we first find the appropriate asymptotic expansions of the resistive solutions, then we match and combine them into a composite solution, which is approximately valid in both regions.\\
The boundary conditions assumed for the unknowns in our problem are the following:
\begin{equation} \label{e:obc}
A_1(1)=B_{\mathrm{re}}, \qquad \psi_1(1)=v_{\mathrm{re}},
\end{equation}
\begin{equation} \label{e:ibc1}
A_1(r_{\mathrm{c}})=0, \qquad \psi_1(r_{\mathrm{c}})=0,
\end{equation}
\begin{equation} \label{e:ibc0}
{A_0}^{\prime}(r_{\mathrm{c}})=0,
\end{equation}
where $B_{\mathrm{re}}$ and $v_{\mathrm{re}}$ are some values of radial components of magnetic and velocity fields, respectively. In the outer region ($|r-r_{\mathrm{c}}| \gtrsim \sqrt{\eta}$) the resistive terms in (\ref{e:i1}) and (\ref{e:i2}) are small and so the outer solution coincides in the lower order approximation with the ideal solution considered in the previous section. Using boundary conditions (\ref{e:obc}) and the above definition of $r_{\mathrm{c}}$, eq. (\ref{e:A1p1}) can be written as
\begin{equation}   \label{e:A1p1bc}
\psi_1=-\frac{v_{\mathrm{re}}}{\ln r_{\mathrm{c}}}\ln \left(\frac{r}{r_{\mathrm{c}}}\right), \qquad A_1=-\frac{B_{\mathrm{re}}}{\ln r_{\mathrm{c}}}\ln \left(\frac{r}{r_{\mathrm{c}}}\right). 
\end{equation}
In this form, the outer solution (\ref{e:A1p1bc}) satisfies the inner boundary conditions (\ref{e:ibc1}) and so it can be used as an inner solution. A similar fact has been observed by Priest et al. \cite{Pr00} in the ``Cartesian'' analogue of our solution. The problem thus reduces to find the inner solutions for ${A_0}^{\prime}$ and ${\psi_0}^{\prime}$. We assume that the following expansions 
\begin{equation} 
{A_0}^{\prime}={A^{0}_0}^{\prime}+\sqrt{\eta}{A^{1}_0}^{\prime}, \qquad {\psi_0}^{\prime}={\psi^{0}_0}^{\prime}+\sqrt{\eta}{\psi^{1}_0}^{\prime},
\end{equation}
are valid in the inner region:
These expressions are inserted into eq. (\ref{e:i3}) and (\ref{e:i2}) where $A_1$ and $\psi_1$ are replaced by the first three terms of the series expansions of (\ref{e:A1p1bc}) about $r_{\mathrm{c}}$. The two resulting equations are then rewritten in terms of the inner variable 
\begin{displaymath}
s=\frac{r-r_{\mathrm{c}}}{\sqrt{2 \eta}}.
\end{displaymath}
The terms with the same powers of $\eta$ are gathered in these equations to equate separately to zero the corresponding coefficients of the two lowest powers of $\eta$. This yields us four equations for the unknowns ${A^{0}_0}^{\prime}$, ${A^{1}_0}^{\prime}$, ${\psi^{0}_0}^{\prime}$ and ${\psi^{1}_0}^{\prime}$. Their solution, rewritten in terms of the variable $r$, determine approximate expressions for ${A_0}^{\prime}$ and ${\psi_0}^{\prime}$ in the vicinity of $r_{\mathrm{c}}$. The latter is expanded then by small $\eta$ at a fixed $r$ and matched with the series expansion of the outer solution about $r=r_{\mathrm{c}}$. Finally, the matched outer and inner solutions are combined into the following composite solution:
\begin{eqnarray}
&{A_0}^{\prime} = E \left[ {\rm{daw}} (\sqrt{k}s) \left(-\frac{2}{3}\frac{\sqrt{k}s^3}{r_{\mathrm{c}}} +\frac{\sqrt{2}}{r_{\mathrm{c}} \sqrt{k}}s^2+  \frac{2}{\sqrt{k}r_{\mathrm{c}}}s+\frac{1}{2r_{\mathrm{c}} k}-\frac{1}{\sqrt{\eta}}\sqrt{\frac{2}{k}}\right) \right.  \nonumber \\
& \left. +\frac{1}{{r_{\mathrm{c}}k}}\left(\frac{7}{3}\mathrm{e}^{-ks^2}+\frac{k}{3}s^2 -\frac{s}{\sqrt{2}}-\frac{5}{6}\right)+\frac{1}{k\sqrt{2\eta}s}+\frac{s\sqrt{2\eta}+r_{\mathrm{c}}}{k{r_{\mathrm{c}}}^2\ln\left(\frac{r_{\mathrm{c}}}{s\sqrt{2\eta}+r_{\mathrm{c}}}\right)}\right] \nonumber \\
& -\left(a{r_{\mathrm{c}}}+\frac{b}{{r_{\mathrm{c}}}}\right)\frac{B_{\mathrm{re}}}{v_{\mathrm{re}}}\mathrm{e}^{-ks^2}+\left[a(s\sqrt{2\eta}+{r_{\mathrm{c}}})+\frac{b}{s\sqrt{2\eta}+r_{\mathrm{c}}}\right]\frac{B_{\mathrm{re}}}{v_{\mathrm{re}}} \nonumber \,
\end{eqnarray}
\begin{equation}
{\psi_0}^{\prime}=\frac{B_{\mathrm{re}}}{v_{\mathrm{re}}}{A_0}^{\prime}+\left[1-{\left(\frac{B_{\mathrm{re}}}{v_{\mathrm{re}}}\right)}^2\right]\left( a{\it r_{\mathrm{c}}}+{\frac {b}{{\it r_{\mathrm{c}}}}} \right), \nonumber
\end{equation}
where $k=({v_{\mathrm{re}}}^2-{B_{\mathrm{re}}}^2)/(v_{\mathrm{re}}{r_{\mathrm{c}}}^2 \ln(r_{\mathrm{c}}))$ must be positive to avoid unphysical divergence of the solution at $\eta\rightarrow 0$. 

\begin{figure} 
\begin{center}
\includegraphics[width=12cm]{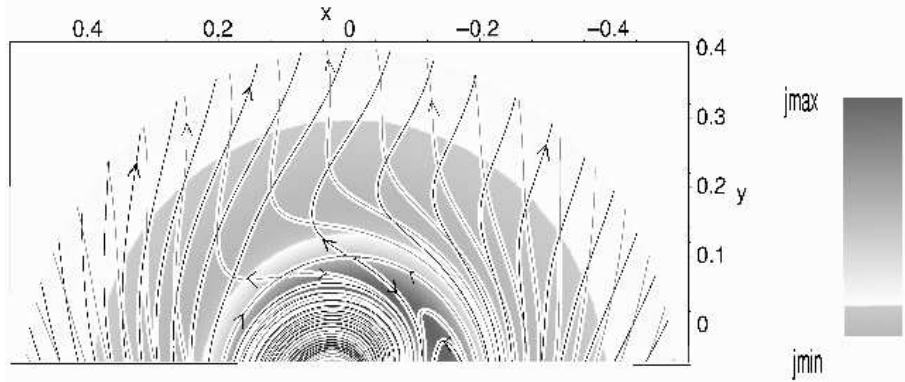}
\end{center}
\caption{Magnetic field lines (solid) and streamlines (dotted) for $d=0.05$, $E=0.5$, $\eta=10^{-2}$, $v_{\mathrm{re}}=0.8$, $B_{\mathrm{re}}=0.9$, $r_{\mathrm{c}}=0.2$, $a=0$ and $b=1$. The plots of the field lines are superimposed to the distribution of the current density in gray half-tones.}
\label{fig:magvel}
\end{figure}

The corresponding magnetic and velocity field lines are plotted in Fig. \ref{fig:magvel}. One can see from this plot that the separatrix line $r=r_{\mathrm{c}}$ is the same for both sets of lines, while the other separatrices are different. The separatrices intersect at the magnetic null point $(r=r_{\mathrm{c}}, \theta=0)$ and at the stagnation point $(r=r_{\mathrm{c}}, \theta=(ar_{\mathrm{c}}+b/r_{\mathrm{c}})k{r_{\mathrm{c}}}^3 {\ln r_{\mathrm{c}}}^2/{v_{\mathrm{re}}}^2)$. Such a structure implies the presence of a shearing component of the flow parallel to the first separatrix $r=r_{\mathrm{c}}$ and transverse to the second magnetic separatrix. As mentioned above, this is a reason for the current layer formation along the first separatrix, which is confirmed by the corresponding distribution of current density in Fig. \ref{fig:magvel}.\\
In comparison with its ``Cartesian'' analogue \cite{Pr00} the obtained solution is much less symmetric. It describes the plasma flow in curvilinear magnetic configuration with an arc-like current layer separating a dipole-like structure from the surrounding unipolar field. This is particularly interesting for modeling magnetic reconnection in solar flares. In fact there are observational evidences that many flares occur in configurations with three magnetic flux concentrations on the photosphere \cite{Ni97}. The proof of this conjecture as well as the detailed investigations of other solutions with $\lambda \neq 0$ will be presented in a forthcoming paper.

\section{Conclusions}
Solutions to the steady incompressible resistive magnetohydrodynamics equations in a curvilinear geometry are derived and discussed. These solutions describe a process where a sheared flow crosses a separatrix of the magnetic field and a curved current layer is formed in correspondence to the other separatrix. These solutions are the analogous in polar coordinates of the solutions in Cartesian coordinates discussed in \cite{Cr95} but with respect to the latter they present some new feature. In particular the configurations of the magnetic and velocity field in the curved geometry is much less symmetric than the one described by the solutions in Cartesian coordinates. Finally the possible applications of our solutions to the modeling of solar flares are mentioned.  

\begin{ack}
The authors would like to gratefully acknowledge the financial support from the Volkswagen-Foundation and from the E.U. Research Training Network HPRN-CT-2000-00153. 
\end{ack}

\bibliographystyle{plain}

\begin{thebibliography}{10}
\bibitem[1]{PrFbs00} 
Priest E.R. and Forbes T.G. 2000
{\em Magnetic reconnection} Cambridge University Press.
\bibitem[2]{Bi00}
Biskamp D. 2000
{\em Magnetic reconnection in plasmas} Cambridge University Press.
\bibitem[3]{Pa57}
Parker E.N . 1957
{\em J. Geophys. Res.} 62, 509-520.
\bibitem[4]{Sw58}
Sweet P.A. 1958
{\em IAU Symp.} 6, 123-134.
\bibitem[5]{Pe64}
Petschek H.E. 1964
{\em Physics of solar flares} 425-439. NASA Report SP-50, Washington DC.
\bibitem[6]{So75}
Sonnerup B.U.O. and Priest E.R. 1975
{\em J. Plasma Phys.} 14, 283-294.
\bibitem[7]{Cr95}
Craig I.J.D. and Henton S.M. 1995
{\em Astrophys. J.} 450, 280-288.
\bibitem[8]{Pr00}
Priest E.R., Titov V.S., Grundy R.E. and Hood A.W. 2000 
{\em Proc. R. Soc. Lond. A} 456, 1821-1849.
\bibitem[9]{Pr96}
Priest E.R. and Titov V.S. 1996
{\em Phil. Trans. R. Soc. Lond. A} 354, 2951-2992.
\bibitem[10]{Ni97}
Nishio M., Yaji K., Kosugi T., Nakajima H. and Sakurai T. 1997
{\em Astrophys. J.} 489, 976-991.

\end{thebibliography}

\end{document}